\documentclass[aps,prb,twocolumn,superscriptaddressgroupaddress,floatfix]{revtex4-2}
\usepackage{amssymb}
\usepackage{physics}
\usepackage{graphicx}
\usepackage{times}
\usepackage{amsmath}
\usepackage{amsthm}
\usepackage{amsfonts}
\usepackage[T1]{fontenc}
\usepackage[utf8]{inputenc}
\usepackage{array}
\usepackage{multirow}
\usepackage{color}
\usepackage{bm}
\usepackage{color}

\usepackage{bbm}
\usepackage[colorlinks]{hyperref}
\usepackage{babel}
\usepackage{float}
\allowdisplaybreaks[4]
\begin{document}
	\title{Interacting quasiperiodic spin chains in the prethermal regime}
	\author{Yi-Ting Tu}
	\email{yttu@umd.edu}

	\author{David M.\ Long}

	\author{Sankar Das Sarma}

	\affiliation{Condensed Matter Theory Center and Joint Quantum Institute, Department of Physics, University of Maryland, College Park, Maryland 20742, USA}

	\begin{abstract}
		Recent progress in the study of many-body localization (MBL) in strongly disordered interacting spin chains has emphasized the importance of distinguishing finite time prethermal behavior from long time and large volume asymptotics.
		We re-examine a reported non-ergodic extended (NEE) regime in the interacting quasiperiodic Ganeshan-Pixley-Das Sarma model from this perspective, and propose that this regime is a prethermal feature.
		Indeed, we argue that the NEE regime may be identified through a change in the functional form of spin-spin autocorrelation functions, demonstrating that the NEE regime is distinguishable within intermediate-time dynamics.
		This is in contrast with existing conjectures relating the NEE regime to the presence of an asymptotic mobility edge in the single-particle spectrum.
		Thus, we propose a mechanism for the formation of an NEE regime which does not rely on asymptotic properties of the spin chain.
		Namely, we propose that the NEE regime emerges due to regularly spaced deep wells in the disorder potential.
		The highly detuned sites suppress spin transport across the system, effectively cutting the chain, and producing a separation of time scales between the spreading of different operators.
		To support this proposal, we show that the NEE phenomenology also occurs in random models with deep wells but with no mobility edges, and does not occur in quasiperiodic models with mobility edges but with no deep wells.
		Our results support the broad conclusion that there is not a sharp distinction between the dynamics of quasiperiodically and randomly disordered systems in the prethermal regime.
		More specifically, we find that generic interacting quasiperiodic models do not have stable intermediate dynamical phases arising from their single-particle mobility edges, and that NEE phenomenology in such models is transient.
	\end{abstract}

	\maketitle

\section{Introduction}

	Improved understanding of the instabilities of many-body localization (MBL) with random disorder~\cite{Anderson1958,Gornyi2005,Basko2006,Oganesyan2007_PRB,Znidaric2008,Pal2010,Devakul2015,Imbrie2016,Imbrie2016b,DeRoeck2017} have lead to substantial revision of the accepted phase diagram for strongly disordered spin chains~\cite{DeRoeck2017,Thiery2018,Suntajs2020,Schulz2020,Sierant2020thouless,Crowley2020,Crowley2020constructive,Vidmar2021,Sels2021obstruction,Morningstar2022,Sels2022,Peacock2023,Long2023phenomenology,Sierant2024review}. There is no longer consensus regarding the presence of an MBL critical point at finite disorder strength~\cite{Sierant2024review}, and it is clear that if such a critical point does exist, it is at a much larger value of the disorder strength than estimated prior to the last five years~\cite{Morningstar2022,Sels2022}. One of the key lessons from this recent body of work is the important distinction between the asymptotic (large volume and long time) behavior of an eventually-thermalizing disordered chain and its \emph{prethermal} behavior at finite---but potentially extremely long---time scales~\cite{Morningstar2022,Long2023phenomenology}.

	\begin{figure}[b]

		\includegraphics[trim=0 0 0 0, clip, scale=0.8]{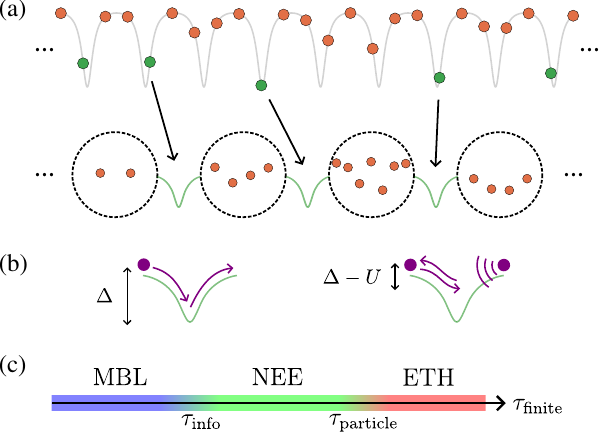}

		\caption{Illustration of the non-ergodic extended (NEE) mechanism in a fermionic chain. (a)~The GPD potential has deep wells (green dots) between typical segments (orange dots). The separating site between the segments is highly detuned, so the chain can be approximated as a similar system where each segment becomes a supersite (dashed circle). (b)~Due to the repulsive interaction, the effective interaction between two supersites (right) is stronger than the hopping between them (left), making the particle spreading timescale $\tau_\text{particle}$ longer than the information spreading timescale $\tau_\text{info}$. (c)~In finite-size or finite-time simulations with maximum accessible timescale $\tau_{\text{finite}}$, the system echibits MBL, NEE, or ETH behavior depending on the relationship between the three time scales. }

		\label{fig:fig1}

	\end{figure}

	Much of the recent progress regarding the asymptotic behavior of strongly disordered chains comes from analyzing the effect of rare regions of low disorder~\cite{Gopalakrishnan2015,Agarwal2015,DeRoeck2017,Morningstar2022,Sels2022}. In quasiperiodically disordered chains, such rare regions do not occur. However, this does not preclude the existence of some other, still unidentified, instability. While heuristic arguments have been made for the stability of asymptotic quasiperiodic MBL in one and two dimensions~\cite{Agrawal2022,Tu2023}, and renormalization schemes predict critical exponents for the MBL transition distinct from the random case~\cite{Khemani2017_PRL,Morningstar2019,Agrawal2020,Tu2023}, the more detailed mathematical analysis of Ref.~\cite{Imbrie2016b} cannot be straightforwardly generalized to the quasiperiodic case.

	On the other hand, it is not obvious that the prethermal behavior of quasiperiodically disordered chains should be distinct from random chains. Very generally, at small system sizes, it should be very difficult to distinguish whether a sequence of \(L \approx 20\) local disorder potentials come from a quasiperiodic sequence or a random one. Nonetheless, several numerical studies at small system size have identified unexpected features in interacting models with quasiperiodic disorder, which have not been previously identified in random models~\cite{Li2015a,Setiawan2017,Hsu2018_PRL,Li2016,Ghosh2020transport,Tu2023b}. In particular, at intermediate disorder strength in the interacting Ganeshan-Pixley-Das Sarma (GPD) model~\cite{Ganeshan2015}, Ref.~\cite{Li2015a} found that the crossover of the half-chain entanglement entropy from being area-law to being volume-law happens at a different point than the crossover of the fluctuation of an observable from being large to being small. This suggest that in the intermediate energy regime, eigenstates are extended (volume-law entangled) while eigenstate expectation values of local operators vary wildly for consecutive eigenstates, failing to satisfy the eigenstate thermalization hypothesis (ETH)~\cite{Deutsch1991,Srednicki1994,DAlessio2016}. This has been identified as a non-ergodic extended (NEE) regime. The same regime in the same model has also been found later by training a neural-network on the half-chain entanglement spectrum to distinguish the ETH, NEE and MBL behaviors~\cite{Hsu2018_PRL}, and by fitting the entanglement entropy growth when coupling the system to a bath to distinguish the three behaviors~\cite{Tu2023b}.

	Note that the term NEE in this paper refers to the failure of the eigenstate thermalization hypothesis while the real-space entanglement entropy satisfies the volume law. In particular, we refer to the extendedness in real space, which is different from the extendedness in Fock space~\cite{Tikhonov2021,Kravtsov2015}, a different topic. 

	Previous work has speculated that the existence of the NEE regime could be related to the presence of a mobility edge in the single particle spectrum of the GPD model~\cite{Li2015a,Li2016}, by explaning the NEE states as the result of interaction mixing two types of non-interacting energy eigenstates: ones that fill only localized single-particle orbitals, and ones that fill some extended orbitals.  The existence of this regime has also been described as a signature of an asymptotic many-body mobility edge (MBME)~\cite{Basko2006,Luitz2015,Li2015a,Li2016,Chanda2020}. (Though, note that there are analytical arguments against the existence of asymptotic MBMEs~\cite{DeRoeck2016,Huang2023b}.)

	However, mobility edges are an asymptotic feature, and numerics of the kind reported in Refs.~\cite{Li2015a,Hsu2018_PRL,Li2016,Ghosh2020transport,Tu2023b} are necessarily restricted to the prethermal regime. In this article, we re-examine several quasiperiodically disordered models, including the GPD model, from the perspective of the prethermal regime. In particular, in Sec.~\ref{sec:AA_GPD}, we analyze the decay of autocorrelation functions, and compare their behavior to the predictions of the statistical Jacobi approximation (SJA)---a technique developed in Ref.~\cite{Long2023phenomenology} (also see Ref.~\cite{Long2023SJA}) to relate the statistics of many-body resonances~\cite{Gopalakrishnan2015,Khemani2017critical,Villalonga2020eigenstates,Crowley2020constructive,Garratt2021,Garratt2022} to autocorrelation functions in the prethermal regime (Sec.~\ref{sec:SJA}). The SJA accurately accounts for the behavior of the interacting Aubry-Andr\'e (AA) model~\cite{Harper1955,Aubry1980}, indicating that its thermalization at intermediate disorder is controlled by a proliferation of successive resonances, much like random models. However, the GPD model shows a regime---qualitatively coinciding with previous estimates for the position of the NEE regime---where the SJA fails to be predictive, and autocorrelators decay with a distinct functional form compared to the AA model and known random models. We interpret this as a dynamical manifestation of the NEE behavior.

	Based on the appearance of NEE signatures well within the prethermal regime, we propose a mechanism for the formation of an NEE regime which does not rely on large system sizes or long times (Sec.~\ref{sec:NEE}). Namely, we observe that the GPD potential has approximately regularly spaced deep wells which may effectively cut the chain (Fig.~\ref{fig:fig1}), and demonstrate in a toy model that this can lead to a separation of timescales between the spreading of different operators. For instance, magnetization may spread much slower than information in general spreads. At small system size, this means that eigenstates will be volume law entangled, to support the spread of information across the whole system, but the expectation value of, say, the spin projection \(S^z_i\) in different eigenstates can vary greatly, as this operator need not have thermalized. This is precisely the phenomenology originally used to identify the NEE regime~\cite{Li2015a}.

	To validate our explanation, we further investigate two more models (Sec.~\ref{sec:t1t2_randomwell})---the \(t_1\)--\(t_2\) model, which has a single-particle mobility edge~\cite{Huang2023}, and a model with random disorder and regularly spaced deep wells, the single-particle sector of which is localized at all energies. We refer to this as the \emph{random wells} model. In this case, we see that the quasiperiodic \(t_1\)--\(t_2\) model shows no NEE regime, while the random wells model does, demonstrating that the NEE regime is unrelated to the asymptotic behavior of the single particle sector, and, indeed, does not rely on quasiperiodicity. Our results support the broad conclusion that there is not a sharp distinction between the prethermal dynamics of quasiperiodically and randomly disordered spin chains.

	Historically, the observation of the NEE regime~\cite{Li2015a,Hsu2018_PRL,Li2016,Ghosh2020transport,Tu2023b} (interpreted as an NEE phase) in the interacting GPD model~\cite{Ganeshan2015} hinted at a possible role of the SPME in dictating  interacting dynamics. This was supported by the observation of three different dynamical regimes (MBL, NEE, ETH) in the interacting GPD model, in contrast to just the two dynamical regimes (MBL and ETH) observed in the extensively studied interacting Anderson (random) model and interacting AA model, which have no SPMEs. The putative existence of an NEE phase (or equivalently an MBME associated with the NEE transition) was even reported in an experiment on a synthetically prepared interacting GPD model in cold atoms~\cite{An2021}. (We note in this context that the GPD model has often been called the generalized Aubry-Andr\'e (GAA) model in the literature. We find this nomenclature to be misleading since there are many possible generalizations of the Aubry-Andr\'e model~\cite{Biddle2010,Huang2023,Wang2020}, which have very different properties compared to the GPD model.) By contrast, cold atom experiments on a different generalization of the AA model, closely related to the $t_1$--$t_2$ model, did not decisively find any NEE phase in the interacting case~\cite{Kohlert2019}, although the corresponding noninteracting system showed the SPME decisively~\cite{Luschen2018}. Our current work takes on particular significance in this context, establishing the putative NEE phase of the interacting GPD model to be a prethermal ``effective phase'' (that is, long lasting transient), which arises from the specific details of the GPD model (its deep potential wells) and not from any SPME or even quasiperiodicity in the system.  The fact that we find NEE features in the random wells model, but do not find NEE features in the quasiperiodic $t_1$--$t_2$ model clearly shows that the observed NEE behavior is a peculiar---though highly interesting---feature of the interacting GPD model.  Our work also conceptually combines all earlier work on the GPD model into one unified scenario:  the interacting GPD model has nontrivial prethermal dynamics involving widely different time scales, which produce the effective NEE behavior, which is, however, a prethermal transient rather than anything connected with the existence of a SPME in the noninteracting GPD model.

	The rest of this article is organized as follows. In Sec.~\ref{sec:SJA}, we briefly review the statistical Jacobi approximation (SJA); in Sec.~\ref{sec:AA_GPD} we study the interacting AA and GPD models; in Sec.~\ref{sec:NEE} we demonstrate that the NEE regime may arise due to the presence of deep wells in the GPD model; in Sec.~\ref{sec:t1t2_randomwell} we study the quasiperiodic $t_1$--$t_2$ model and the random wells model in the context of the NEE behavior; we conclude in Sec.~\ref{sec:conclusion} with a summary and discussions of future directions of research.  Two appendices complement the results in the main text by providing the eigenstate properties of the random wells model (Appendix~\ref{sec:randw_eigen}) and the GPD model with the next-nearest-neighbor hopping (Appendix~\ref{sec:t1t2GPD}) as a verification of our proposal.

	\section{The statistical Jacobi approximation}
	\label{sec:SJA}

	It has been proposed that the thermalization of randomly disordered spin chains at intermediate disorder is controlled by the proliferation of many-body resonances~\cite{Villalonga2020eigenstates,Crowley2020constructive,Schulz2020,Garratt2021,Garratt2022,Long2023phenomenology}. Initial states slowly Rabi oscillate between \emph{resonant} macroscopically distinct spin configurations of almost equal energy. Longer evolution times resolve finer and finer successive resonances, which eventually lead to thermalization. Refs.~\cite{Long2023phenomenology,Long2023SJA} introduced the \emph{statistical Jacobi approximation} (SJA) to relate the proliferation of resonances to the observable behavior of autocorrelation functions. The SJA does not seem to rely on any particular feature of the disorder potential, and Ref.~\cite{Long2023phenomenology} speculated that it would continue to be predictive for spin chains with correlated disorder, including quasiperiodic disorder.

	The prediction of the SJA serves as our null hypothesis in studying a potential non-ergodic extended regime. If the SJA accurately predicts the behavior of autocorrelation functions, then it is likely that thermalization is controlled by many-body resonances, just as in other prethermal models. However, if the SJA fails to be predictive, it is an indication that thermalization proceeds through an alternative mechanism, or potentially that the model is non-ergodic.

	The SJA is based on the Jacobi algorithm~\cite{Jacobi1846}, an iterative algorithm for matrix diagonalization. In addition to obtaining the exact eigenstates of the Hamiltonian, this algorithm provides a way to extract a basis of near-eigenstates associated to a finite time scale---the Jacobi basis. This makes it a useful tool for the prethermal regime, as some control over the statistical properties of the Jacobi basis can be maintained at intermediate times, which allows the prediction of dynamical observables.

	For a Hamiltonian $H$ represented as a matrix in a local tensor product basis, the algorithm starts by finding the off-diagonal element $H_{ab}$ (where $a$ and $b$ label the basis states and $a\neq b$) with the largest absolute value, diagonalizing the block
	\begin{equation}
	  U^\dagger
	  \begin{pmatrix}
		H_{aa} & H_{ab} \\
		H_{ba} & H_{bb}
	  \end{pmatrix}
	  U =
	  \begin{pmatrix}
		H'_{aa} & 0 \\
		0 & H'_{bb}
	  \end{pmatrix}
	\end{equation}
	and extending this to the entire $H$ as
	\begin{equation}
	  H' = (U^\dagger \oplus I)H(U \oplus I).
	\end{equation}
        This procedure is then repeated iteratively (the transformed basis in $H'$ becomes the new basis that $a$ and $b$ label), so that $H$ becomes increasingly diagonal (that is, $\sum_{a\neq b} |H_{ab}|^2$ decreases strictly and converges to zero).
	If at some step $|H_{ab}|$ is much larger than $|H_{aa}-H_{bb}|$, the corresponding basis state at $a$ and $b$ becomes \emph{resonant}. That is, the new basis has
	\begin{equation}
	  |a'\rangle\approx\frac{|a\rangle+|b\rangle}{\sqrt2},\quad |b'\rangle\approx\frac{|a\rangle-|b\rangle}{\sqrt2}.
	\end{equation}

	The quantity of interest in the SJA is the number density of the resonant decimated elements $w=|H_{ab}|$, which we called the \emph{distribution of resonances} $\rho_\text{res}(w)$. Moreover, this can be energy-resolved by targeting a specific energy $E$ and collecting only the decimated elements with either $H_{aa}$ or $H_{bb}$ close to $E$~\cite{Long2023SJA}.
	The distribution of resonances as a function of $w=|H_{ab}|$ can be characterized by a power law at intermediate timescales (that is, intermediate values of $w^{-1}$)~\cite{Long2023phenomenology},
	\begin{equation}
	  \rho_\text{res}(w,E) \propto w^{-1+\theta(E)},
	\end{equation}
	where $\theta(E)$ is called the \emph{resonance exponent}.

	Under some mild assumptions~\cite{Long2023phenomenology}, the SJA then relates $\rho_\text{res}$, and more specifically \(\theta(E)\), to the functional form of autocorrelation functions. In particular, the main observable used in the numerical calculation of this paper, where the system is a one-dimensional (1D) spin-$1/2$ chain, is the energy-resolved spin-spin connected autocorrelator,
	\begin{multline}
	  C(E,t) = \left[\langle\psi_E| S^z_{L/2}(t) S^z_{L/2}(0) |\psi_E\rangle\right]\\
	  - \left[\langle\psi_E| S^z_{L/2}(0) |\psi_E\rangle\right]^2,
	\end{multline}
	where $L$ is the length of the chain, $|\psi_E\rangle$ is a Haar-random superposition of the 100 energy eigenstates with energy closest to $E$, \(S^z_j(t)\) is the Heisenberg picture spin-\(z\) operator on site \(j\) at time \(t\), and square brackets indicates an average over disorder and the states \(|\psi_E\rangle\). The term \([\langle\psi_E| S^z_{L/2}(0) |\psi_E\rangle]\) coincides with the expectation value of \(S^z_{L/2}\) in a microcanonical shell around energy \(E\), with the average over Haar random states serving as a low variance estimator of the trace~\cite{Facchi2015}.

	The SJA predicts that \(C(E,t)\) is described by a stretched exponential at intermediate timescales~\cite{Long2023phenomenology},
	\begin{equation}
	  C(E,t) \approx A(E)\,\exp\left[-\left(\frac{t}{\tau(E)}\right)^{\beta(E)}\right],
	  \label{eqn:stexp_fit}
	\end{equation}
	where $\tau(E)$ is the thermalization timescale and $\beta(E)=-\theta(E)$ is the \emph{stretch exponent}. We numerically calculate $C(E,t)$ by first obtaining the eigenstates closest to $E$ using the polynomially filtered exact diagonalization (POLFED) algorithm~\cite{Sierant2020}, and then time evolving the random superposition $|\psi_E\rangle$, as well as $S_{L/2}^z|\psi_E\rangle$ under the Hamiltonian \(H\) using the Krylov subspace projection method~\cite{Sidje1998,Croy2018}. The disconnected piece of \(C(E,t)\) can then be computed for a particular disorder realization as the matrix element \((\langle \psi_E| e^{i H t}) S^z_{L/2} (e^{-i H t} S^z_{L/2} |\psi_E\rangle) \). The term \(\langle\psi_E| S^z_{L/2}(0) |\psi_E\rangle\) can be computed from the initial conditions.

	After obtaining $C(E,t)$, least-squares fitting is used to estimate $\beta(E)$ and $\tau(E)$.
	However, for a finite size system, $C(E,t)$ will approach a fixed nonzero value at large $t$, and hence we add a constant shift $B(E)$ as an additional phenomenological parameter to fit $C(E,t)$. This allows us to fit the curve using a larger time range, which we observe makes the fit more constrained, compensating for the additional freedom of using an extra fit parameter.

	Either the failure of the stretched exponential fit, unphysical fit parameters, or the deviation of $\beta$ from the value of $-\theta$ computed from the Jacobi algorithm indicates a failure of the SJA. This failure may be an indication that the system is non-ergodic.

	\section{The GPD model}\label{sec:AA_GPD}

	\begin{figure}
		\includegraphics[trim=3 0 0 0, clip, scale=0.8]{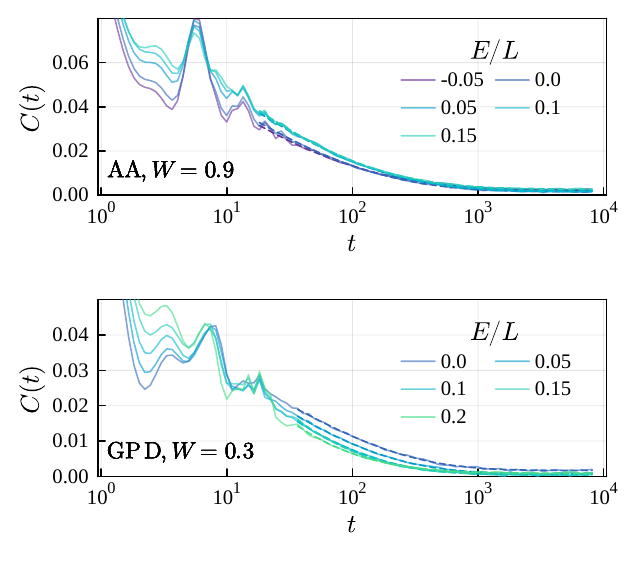}
		\caption{The decay of the spin-spin connected autocorrelation function $C(t)$ of the AA and GPD models (Eqs.~(\ref{eq:Heisenberg},~\ref{eq:GPD}) with \(\alpha=0\) and \(\alpha=-0.8\) respectively) at various energy densities. Dashed curves are the shifted stretched exponential fit of Eq.~\eqref{eqn:stexp_fit}.}
		\label{fig:Ct}
	\end{figure}

	\begin{figure*}
		\includegraphics[trim=0 0 0 0, clip, scale=0.8]{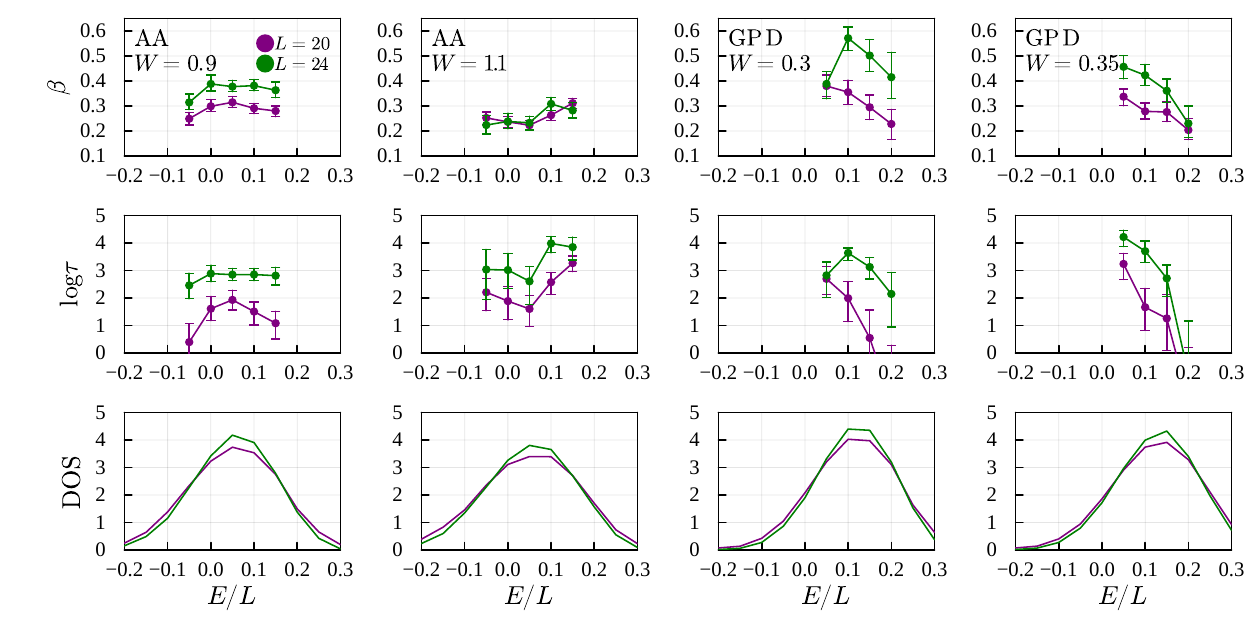}
		\caption{The coefficients $\beta$ (top) and $\log\tau$ (middle) obtained from the shifted stretched exponential fit of $C(t)$, and the normalized density of states (bottom) for the AA (left) and GPD (right) models at two selected values of $W$, each as a function of energy density. The GPD model shows a sudden drop to $\tau\to 0$ at high energy density (especially for $L=20$). Error bars indicate the $68\%$ bootstrap confidence interval due to the random choice of $\phi$. }
		\label{fig:fits}
	\end{figure*}

	\begin{figure}
		\includegraphics[trim=0 0 0 0, clip, scale=1]{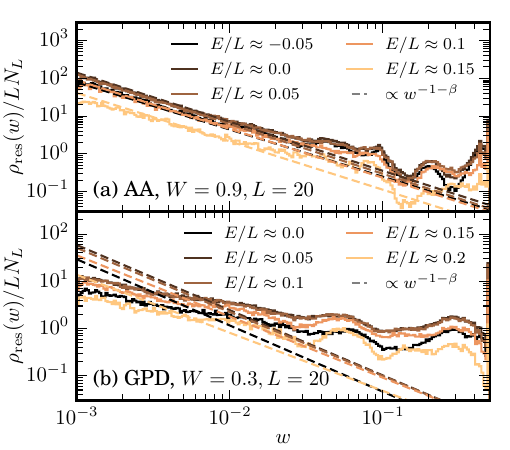}
                \caption{The distribution of resonances $\rho_\text{res}$ obtained from the Jacobi algorithm and the comparison with the predicted power law from the stretched exponent $\beta$. These agree well for the AA model but not for the GPD model.}
		\label{fig:rhores_beta2}
	\end{figure}

	The interacting GPD model~\cite{Ganeshan2015} is a spin-$1/2$ Heisenberg model
	\begin{equation}\label{eq:Heisenberg}
		H = \sum_{j=1}^{L-1} \left( S_j^x S_{j+1}^x + S_j^y S_{j+1}^y + S_j^z S_{j+1}^z \right) + W\sum_{j=1}^{L} h_j S_j^z,
	\end{equation}
	with on-site quasiperiodic potential
	\begin{equation}\label{eq:GPD}
		h_j = \frac{\cos(2\pi\varphi j + \phi)}{1-\alpha\cos(2\pi\varphi j + \phi)}.
	\end{equation}
	Here, $S_{j}^{x,y,z}$ are the spin-$1/2$ operators at site $j$ along the $x,y,z$ axes, $W$ is the disorder strength, the spin exchange amplitude has been set to 1, $\varphi=\frac{1+\sqrt{5}}{2}$ is the golden ratio, $\phi$ is an initial phase (which are averaged over \(50\) to \(400\) random choices), and $\alpha\in(-1,1)$ is a dimensionless parameter. When $\alpha=0$, Eq.~\eqref{eq:Heisenberg} reduces to the interacting Aubry-Andr\'e (AA) model~\cite{Harper1955,Aubry1980}. When $\alpha<0$ ($>0$), the cosine potential of the AA model is distorted so that the peaks become sharper at negative (positive) energy. We focus on the case of $\alpha=0$ (AA) and $-0.8$ (GPD), where in the latter case the sharp peaks at negative energy cause some $h_{j}$ to become \emph{deep wells} which are highly detuned from their neighboring sites (Fig.~\ref{fig:fig1}a). We fix the filling fraction to $1/4$ (that is, $1/4$ of the spins are up), since the proposed effect of an MBME would be more distinguished at a lower filling fraction, and $1/4$ is still high enough to make the effect of interaction significant.

	In Fig.~\ref{fig:Ct}, we plot $C(t)$ for several energy densities $E/L$, as well as the shifted stretched exponential fit Eq.~\eqref{eqn:stexp_fit} for the AA and GPD models. The fit parameters, and also the density of states (DOS) at that $E/L$, are shown in Fig.~\ref{fig:fits}. We find that the AA model shows the expected stretched exponential behavior as predicted by the SJA. However, at higher energy densities (\(E/L \gtrsim 0.15\)) of the GPD model, we observe a sudden drop of the characteristic timescale $\tau$ toward zero, indicating $C(t)$ becomes essentially scale-invariant, which is not consistent with the stretched exponential prediction. The value of $\beta$ also decreases.

	This observation holds across the entire range of $W$ where it is sensible to make the fit at the available timescale. That is, where $C(t)$ slowly decays to a value substantially smaller than its initial value. For smaller or larger $W$, the decay is either too fast or too slow to observe the functional form of the decrease in $C(t)$, such that a meaningful and stable fit is impossible. Also note that we only show the data for energy densities such that the normalized density of states (DOS) is larger than $2$ (roughly the middle two-thirds of the spectrum), since the fit at the edge of the spectrum is again very unstable. This crossover occurs near the peak of the DOS, rather than at the edge of the spectrum.

	However, we do not expect this phenomenon to survive the thermodynamic limit. This can be seen in the comparison between $L=20$ and $L=24$ in Fig.~\ref{fig:fits}, where the sudden drop of $\tau$ appears to be less dramatic and happens at slightly larger $E/L$ for the $L=24$ curves. This means that the stretched exponential fit, and therefore the SJA, is better at larger $L$. Therefore, we suspect that the failure of SJA is a finite-size effect.

	We also compare the stretch exponent $\beta$ to the resonance exponent $\theta$ obtained directly by running the Jacobi algorithm. The result is shown in Fig.~\ref{fig:rhores_beta2}. The distribution of resonances $\rho_\text{res}(w)$ shows power-law behavior except for very large $w$, which is dominated by short-time effects. The expected power law curves from the stretched exponent are shown in dashed lines. For the AA model, the two sets of curves agree very well. However, for the GPD model, they do not agree for some values of $W$. This indicates that the AA model agrees with the model of resonances even at fixed energy density, but the GPD model does not.

	These results suggest that thermalization in the interacting GPD model is not described well by the proliferation of many-body resonances. In particular, the failure of the SJA in the higher energy part of the model may indicate that the system is non-ergodic on these time scales.
	% Also note that the actual decay of $C(t)$ becomes faster compared to lower energy density (see Fig.~\ref{fig:Ct}), suggesting that the higher energy part is more extended.
	Moreover, the position of this regime agrees qualitatively with previous proposals for the position of the NEE regime in the GPD model~\cite{Li2015a,Hsu2018_PRL,Li2016,Tu2023b,YiTing_ML}. (Note that the regimes obtained by the studies based on finite-time dynamical simulation are expected to differ quantitatively from those based on eigenstate properties.) Therefore, we interpret the regime of the GPD model where the stretched exponential fit fails as the NEE regime, and conclude that thermalization in this regime proceeds through a different mechanism than the AA model and typically studied random models.
	Note that the indication of the NEE behavior through the finie-time SJA study suggests that the NEE regime \emph{can} appear as a prethermal feature. Since it is logically possible that there exist an asymptotic NEE phase that cause the failure of SJA at all timescales, including the timescale at which we study, our observation above do not logically imply that the NEE behavior is \emph{only} a prethermal feature. However, we believe that the latter is true, as will be explained and discussed in the next two sections.

	\section{Origin of the NEE behavior}\label{sec:NEE}

	The NEE regime has been conjectured to be related to the presence of a single-particle mobility edge (SPME) in the GPD model~\cite{Li2015a,Li2016}. The AA model has no energy-dependent mobility edge, and does not exhibit an NEE regime. However, we observe that the NEE behavior manifests in the functional form of autocorrelation functions at intermediate timescales and small system sizes, where the influence of a mobility edge in the asymptotic single particle spectrum would be expected to be minimal. In addition, the drift with $L$ of the crossover position suggests that the regime will not survive the thermodynamic limit, and hence is not due to a many-body mobility edge (MBME) (and there are theoretical arguments against the stability of an MBME~\cite{DeRoeck2016,Huang2023b}).
	Instead, we propose that the NEE behavior comes from the presence of deep wells in the GPD potential. The large detuning effectively cuts the chain into several segments, with spin exchange being strongly suppressed between segments, but other interactions being less suppressed.

	In this section, we first construct a toy model with the deep well structure whose NEE behavior can be derived analytically using perturbation theory, and discuss the relationship between the toy model and the actual GPD model. In the next section, we provide further numerical evidence to support this theory.

	For convenience, we formulate the toy model Hamiltonian in fermionic notation. It takes the form of a Hubbard model
	\begin{equation}
	  H=\sum_{j}\left(J\,(c_j^\dagger c_{j+1}+\text{H.c.}) + U n_j n_{j+1} + V_j n_j\right),
	\end{equation}
	where $c_j$ is the fermion annihilation operator and $n_j=c_j^\dagger c_j$ is the fermion number operator. The on-site potential consists of random variables $V_j \in [\Delta_j-\mathcal{W},\Delta_j+\mathcal{W}]$, where
	\begin{equation}
	  \Delta_j =
	  \begin{cases}
		0 & \text{if }j\neq 0\mod N+1,\\
		- \Delta & \text{if }j=0\mod N+1.\\
	  \end{cases}
	\end{equation}
	Here, $N$ is the size of the \emph{supersites} which we introduce below. We take the intra-supersite disorder width $\mathcal{W}\ll J$, and $\Delta\gg J$, so that there is a large detuning (deep well) every $N+1$ sites that suppresses the particle from hopping across the deep well. On the other hand, we choose the scale of the repulsive interaction $U$ to satisfy $\Delta \gg \Delta-U\gg J$. We show below that this leads to a parametric separation of the particle spreading and information (and, indeed, energy) spreading timescales.

	We construct an effective Hamiltonian for the middle-energy subspace (not involving the scales $-\Delta$ or $U$) treating $J$ as a small parameter. This effective Hamiltonian describes energy densities corresponding to the typical sites. The middle-energy subspace is spanned by the particle configurations in which every deep site $j=0\mod N+1$ is unoccupied, and such that no two particles are occupying consecutive sites. In particular, the deep wells effectively cut our chain into segments, which we call the \emph{supersites}, each of size $N$. The leading inter-supersite terms in the effective Hamiltonian can be calculated from second-order perturbation theory. Let $j=0\mod N+1$ be a deep well. The inter-supersite hopping is described by a particle hopping from $j-1$ to $j+1$ (and its reverse), through the process (left part of Fig.~\ref{fig:fig1}b)
	\begin{equation}
	  |\cdots100\cdots\rangle \mapsto |\cdots010\cdots\rangle \mapsto |\cdots001\cdots\rangle,
	\end{equation}
	where the three numbers in each ket are the occupation of $j-1$, $j$, and $j+1$, respectively. This gives the term in the effective Hamiltonian:
	\begin{equation}
	  \frac{J^2}{\Delta}\,c_{j+1}^\dagger c_{j-1} + \text{H.c.}.
	\end{equation}
	On the other hand, the inter-supersite interaction is described by that of two particles at $j-1$ and $j+1$, though the processes (right part of Fig.~\ref{fig:fig1}b)
	\begin{equation}
	  |\cdots101\cdots\rangle \mapsto
	  \left\{
	  \begin{array}{l}
		|\cdots011\cdots\rangle \\
		|\cdots110\cdots\rangle
	  \end{array}
	  \right\}
	  \mapsto |\cdots101\cdots\rangle,
	\end{equation}
	which is to be compared with the processes without the other particle,
	\begin{equation}
	  \begin{array}{l}
		|\cdots100\cdots\rangle \mapsto |\cdots010\cdots\rangle \mapsto |\cdots100\cdots\rangle,\\
		|\cdots001\cdots\rangle \mapsto |\cdots010\cdots\rangle \mapsto |\cdots001\cdots\rangle.
	  \end{array}
	\end{equation}
	This gives the term in the effective Hamiltonian:
	\begin{equation}
	  \left(2\,\frac{J^2}{\Delta-U} - 2\,\frac{J^2}{\Delta}\right)\,n_{j+1} n_{j-1}.
	\end{equation}
	Note that $1/\Delta \ll 1/(\Delta-U)$ by our assumption. This indicates that, in the effective model, particle exchange (hopping) between the supersites is suppressed much more strongly than the information exchange (interaction) between them. Suppose the disorder in $V_j$ is weak enough such that the system in the thermodynamic limit eventually thermalizes (satisfying ETH). We can define the time scale $\tau_\text{particle}$ at which particles can spread across the entire system and $\tau_\text{info}$ for information similarly, so that we have $\tau_\text{info}\ll\tau_\text{particle}$.

	Now suppose that we try to use some finite-size or finite-time probe to study the ergodicity and extendedness of the system, so that there is another time scale $\tau_\text{finite}$ which is the largest time scale that we can probe. The observable properties of the system will depend on the relationship between the three scales \(\tau_{\text{finite}}\), \(\tau_{\text{particle}}\), and \(\tau_{\text{info}}\). In the case of $\tau_\text{finite} \ll \tau_\text{particle}$, the effect of particle spreading will not be observed, and there will appear to be a set of conserved quantities. Namely, the number of particles in each supersite. Thus, the system will appear non-ergodic. For example, the system will show large fluctuations of the half-chain density among nearby eigenstates, which was used in Ref.~\cite{Li2015a} to support the existence of an NEE regime. Conversely, there are no such apparent conserved quantities if $\tau_\text{finite} \gg \tau_\text{particle}$ and if there are no still larger time scales, we will see the asymptotic ergodic behavior. On the other hand, if we have $\tau_\text{finite} \ll \tau_\text{info}$, the entanglement entropy will show area law behavior, while for $\tau_\text{finite} \gg \tau_\text{info}$ the entanglement entropy will be volume law---interpreted in Ref.~\cite{Li2015a} as meaning that the system is extended. Therefore, we may numerically observe the behavior of three different regimes in dynamics at accessible system sizes, even though the system thermalizes in the thermodynamic limit (Fig.~\ref{fig:fig1}c). In particular, if $\tau_\text{info} \ll \tau_\text{finite} \ll \tau_\text{particle}$, this finite-size or finite-time probe will show that the system is in the NEE regime.

	We can further transform our toy model to a more familiar form for the simplest case of $N=2$ and the $1/3$ filling sector. The middle-energy Hilbert subspace has a simple basis
	\begin{equation}
	  \{|n_1,1-n_1,0,n_2,1-n_2,0,\ldots\rangle\}_{n_1,n_2,\ldots\in\{0,1\}},
	\end{equation}
	where the first position denotes site $1$. We can map the basis states to $|n_1,n_2,\ldots\rangle$, which is the state space of the Ising model. In this new Hilbert space, the effective Hamiltonian at leading order becomes
	\begin{multline}
	  H_\text{eff} = \sum_i \Bigg(
			J \sigma_i^x
			+ \frac{V_{3i-1}-V_{3i-2}}{2}\, \sigma_i^z\\
			- \frac{J^2}{2(\Delta-U)}\, \sigma_i^z \sigma_{i+1}^z
		\Bigg),
	\end{multline}
	which is a mixed-field random Ising model. Previous numerical evidence strongly supports that this model thermalizes in the thermodynamic limit if the intra-supersite disorder is not too strong~\cite{Kim2014}. The deep well structure can thus simultaneously support an extensive number of almost-conserved local operators (the supersite occupation numbers) and extended eigenstates.

	In the general case of arbitrary $N$ and filling fractions, the effective model will be much more complicated, but it is still likely to thermalize for weak disorder, so that an NEE regime emerges.

	We emphasize that in the NEE regime of our toy model, where the particle hopping across the deep wells may be neglected, the number of particles in each supersite constitutes an \emph{extensive} set of conserved local integrals of motion (LIOMs), even if it is not a \emph{complete} set. Thus, the effective model for the NEE regime is, in fact, many-body localized. The system retains partial memory of generic initial conditions for all time, and particle transport coefficients vanish. Nonetheless, the eigenstates of this model are volume law entangled, and energy can diffuse through the system. While the model is localized, it is not \emph{fully} localized, in the sense of Ref.~\cite{Huse2014}. One could call our picture of the NEE regime one of \emph{partial many-body localization}.

	It is also apparent that thermalization in the toy NEE model has more structure than apparent from the distribution of resonances \(\rho_{\mathrm{res}}(w)\) obtained from the Jacobi algorithm. Some operators spread much more slowly than others, which is information not resolved by the distribution \(\rho_{\mathrm{res}}(w)\). While many-body resonances may play a role in the melting of the almost-conserved quantities of the NEE toy model, this process must be distinguished from the much more rapid spread of other operators.

	\begin{figure}
		\includegraphics[trim=3 0 0 0, clip, scale=0.8]{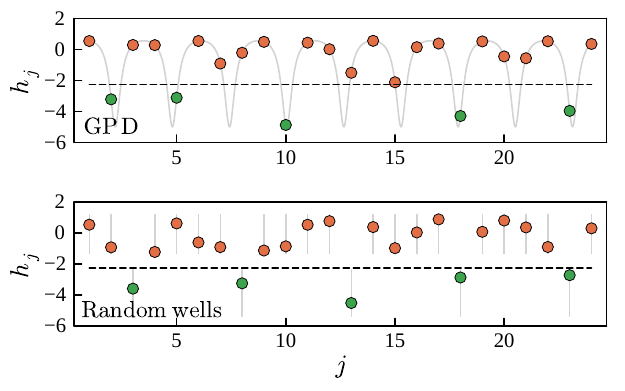}
		\caption{Comparison between the GPD and the random wells model. The gray curve in the top figure is the periodic function whose values at integer points give the GPD potential. We put a horizontal dashed line to separate the deep (green) and shallow (orange) sites. The threshold is chosen such that the number density of deep sites is $1/5$. The random wells model is constructed such that the deep sites are equally spaced with the same number density, and the potentials on the deep and shallow sites are both independently uniformly distributed with the same mean and variance as the corresponding type of sites in the GPD model (the ranges are shown in gray horizontal bars in the bottom figure).}
		\label{fig:randw}
	\end{figure}

	Now we return to the GPD model. The deep well structure happens when the parameter $\alpha$ is close to $-1$ (in our study and the previous numerical evidence~\cite{Li2015a,Hsu2018_PRL,Tu2023b} for the NEE regime in the interacting GPD model, the value $-0.8$ is used, which is close enough), as the sites with $\cos(2\pi\varphi j+\phi)\approx 1$ cause the denominator of Eq.~(\ref{eq:GPD}) to be close to zero. Although there is no canonical choice of how close the cosine term is to one for the site $j$ to be considered as deep, we can still compare with our toy model based on the visual appearance. In the range of disorder strengths used in our numerical calculations, we estimate $\Delta\sim 1.3$, $\mathcal{W}\sim 0.4$, and $N\sim 5$ (see Fig.~\ref{fig:randw}). Also, we have $J=0.5$ and $U=1$ by Jordan-Wigner transformation. We see that the assumption that the detuning of the deep well is compensated by the interaction is indeed satisfied. If we use the perturbative formulae above, we get the inter-supersite hopping strength to be $\sim 0.2$ and the inter-supersite interaction strength to be $\sim 1$, indicating the rates of particle hopping and interaction are separated by a factor of \(\sim 5\), lending support to our proposed mechanism for the NEE behavior. Although perturbation theory may not work well for these parameters, since the energy scales are all roughly at the same order of magnitude, we propose that the essential feature caught by our perturbative toy model is responsible for the observed NEE regime of the interacting GPD model.

	% In the next section, we will further support this by numerically simulating a quasiperiodic model with an SPME but without the deep well structure, and comparing it with a random model (hence without SPME) which does have a deep well structure. We will show that the former does not have an NEE regime, but the latter does.

	\section{The \texorpdfstring{$t_{1}$--$t_{2}$}{t1-t2} model and the random well model}\label{sec:t1t2_randomwell}

	\begin{figure}
		\includegraphics[trim=3 0 0 0, clip, scale=0.8]{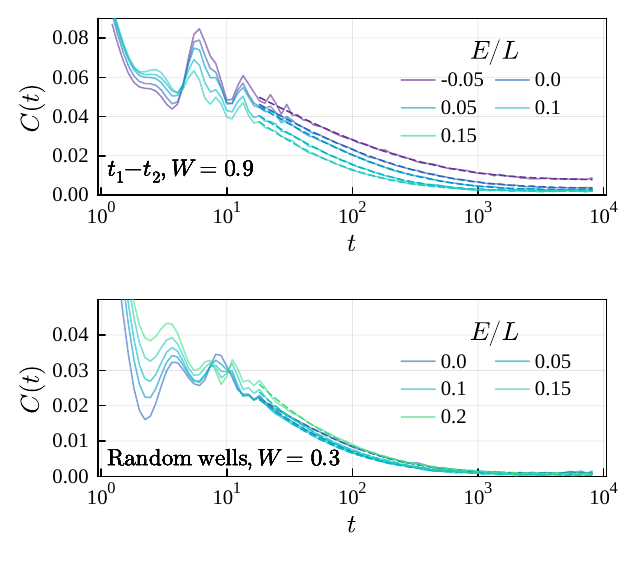}
		\caption{The decay of the spin-spin connected autocorrelation function $C(t)$ of the $t_1$--$t_2$ and the random wells models at various energy densities. Dashed curves are the stretched exponential fit.}
		\label{fig:Ct_2}
	\end{figure}

	\begin{figure*}
		\includegraphics[trim=0 0 0 0, clip, scale=0.8]{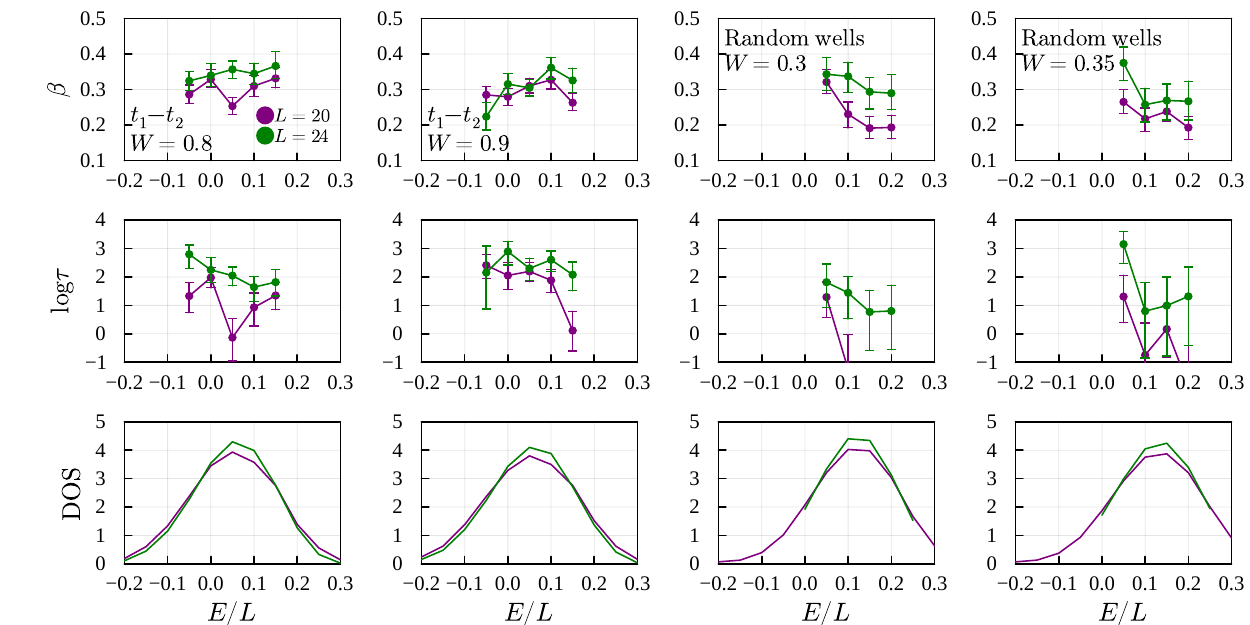}
		\caption{The coefficients $\beta$ (top) and $\log\tau$ (middle) obtained from the shifted stretched exponential fit of $C(t)$, and the density of states (bottom) for the $t_1$--$t_2$ (left) and random wells (right) models at two selected values of $W$, each as a function of energy density. The random wells model shows a sudden drop to $\tau\to 0$ at high energy density (especially for $L=20$). Error bars indicate the $68\%$ bootstrap confidence interval due to the random choice of $\phi$ or disorder realizations. (Compare to Fig.~\ref{fig:fits}.)}
		\label{fig:fits_2}
	\end{figure*}

	We propose that any one-dimensional model with regularly spaced deep wells in its disorder potential will display an NEE regime at intermediate timescales. In particular, our proposal does not rely on the presence or absence of a mobility edge in the single particle spectrum, nor, indeed, on quasiperiodicity. In this section, we investigate two additional models: the \(t_1\)--\(t_2\) model, a quasiperiodic model with a single particle mobility edge but no deep wells, and a randomly disordered model in which we insert deep wells. The single particle sector of the latter is always localized~\cite{Anderson1958}, but we will see that it exhibits an NEE regime, while the \(t_1\)--\(t_2\) model does not.

	The $t_{1}$--$t_{2}$ model is just the AA model with an additional next-nearest-neighbor hopping term~\cite{Luschen2018,Huang2023}.
	In the spin representation, its Hamiltonian is
	\begin{multline}\label{eq:t1t2}
	  H = \sum_{j=1}^{L-1} \left[\frac{1}{2}t_{1}\left( S_j^+ S_{j+1}^- + S_j^- S_{j+1}^+ \right) + U S_j^z S_{j+1}^z \right]\\
	  - \sum_{j=1}^{L-2} t_{2}\left( S_j^+ S_{j+1}^z S_{j+2}^- + S_j^- S_{j+1}^z S_{j+2}^+ \right)\\
	  + W\sum_{j=1}^{L} \cos(2\pi\varphi j+\phi) S_j^z.
	\end{multline}
	Here, $t_{1}$ and $t_{2}$ correspond to the strength of nearest-neighbor and next-nearest-neighbor hopping respectively, and $U$ is the interaction strength. We fix $t_{1}=1$, $t_{2}=1/6$ (as in Ref.~\cite{Huang2023}) and $U=1$.

	The \emph{random wells} model is based on the perturbative toy model in Sec.~\ref{sec:NEE}, with parameters chosen to make the potential visually mimic the GPD model. The Hamiltonian is the same as Eq.~(\ref{eq:Heisenberg}), except that the potentials $h_{j}$ are each independently drawn from a \(j\)-varying uniform random distribution. There is no canonical choice of what points are considered deep in the GPD model, and we do not think the exact definition is important. However, to have some consistency between the GPD and our artificial model, we make an arbitrary choice of threshold $-2.28$, and consider sites with $h_{j}<-2.28$ to be a deep well. This choice makes the number density of deep sites in the GPD model exactly $1/5$. In this way, we can design the artificial model to have a deep well for every five sites, such that the density of deep wells is the same as the GPD model under this definition. The range of the uniform random variable $h_{j}$ is chosen to be $[-5.38, -2.36]$ for $j \bmod 5=j_{0}$ (deep); otherwise it is chosen to be $[-1.33, 1.19]$ (shallow). The values are chosen so that the mean and variance of the potential for a type of site (deep or shallow) is the same as that of the corresponding type in the GPD model. Each of the five possible choices of $j_{0}$ are used in exactly one-fifth of the disorder realizations. The visual comparison between the GPD and our artificial random wells model is shown in Fig.~\ref{fig:randw}.

	As a remark, the eigenstate properties of the random wells model show essentially the same signatures to those of the GPD model that in Ref.~\cite{Li2015a} were used as evidence for the NEE regime (Appendix~\ref{sec:randw_eigen}).

	In Figs.~\ref{fig:Ct_2}--\ref{fig:fits_2}, we plot the $C(t)$ curve and the fit parameters for the $t_1$--$t_2$ and the random wells models, similar to Figs.~\ref{fig:Ct}--\ref{fig:fits}. The results show that the interacting $t_1$--$t_2$ model, despite having a single particle mobility edge, shows similar behavior to the interacting AA model, in that the stretched exponential fit is good (shows sensible parameters). On the other hand, the interacting random wells model, despite being random rather than quasiperiodic, and thus being localized everywhere in the single particle spectrum, shows similar behavior to the interacting GPD model in that the stretched exponential fit crosses over to $\tau\to 0$ behavior in the higher energy spectrum.

	Our proposed mechanism based on deep wells also predicts that adding a next-nearest-neighbor hopping term to a model with deep wells will destroy the NEE regime, as a particle can directly hop from one side of the deep well to the other. Thus, there will be much less suppression of particle spreading, and therefore the NEE regime should disappear. We verify this numerically for the GPD model with next-nearest-neighbor in Appendix~\ref{sec:t1t2GPD}.

	\section{Conclusion}\label{sec:conclusion}

	The asymptotic properties of randomly disordered interacting spin chains are predicted to be distinct in several ways from quasiperiodically disordered spin chains~\cite{Morningstar2019,Agrawal2020,Agrawal2022,Tu2023}. However, at accessible system sizes, few robust differences between the behavior of quasiperiodic and random disorders have been observed~\cite{Khemani2017_PRL}. This indicates that the prethermal regime---in which the system eventually thermalizes, but does so extremely slowly---is phenomenologically the same in random and quasiperiodic spin chains. One of the few observed differences between these cases in past work has been the presence of an apparently non-ergodic extended (NEE) regime in the quasiperiodic GPD model, which has been attributed to the presence of a single-particle mobility edge~\cite{Li2015a,Hsu2018_PRL,Li2016,Tu2023b}.

	By diagnosing the presence of the NEE regime through the functional form of spin-spin autocorrelation functions, we have demonstrated that the NEE phenomenology can be observed within the prethermal regime, and so is unlikely to be due to the asymptotic presence of a single-particle mobility edge. Rather, we propose that this regime emerges due to the presence of deep wells in the GPD potential, which suppress spin exchange across the chain. This proposal is also consistent with the failure of the statistical Jacobi approximation (SJA) to describe the decay of autocorrelators, as the thermalization process is different for different operators, and the SJA does not resolve such differences. Notably, this mechanism does not rely on quasiperiodicity. Indeed, we can replicate the same phenomenology in a randomly disordered model with deep wells. Thus, while our results show the unremarkable feature that details of the disorder potential may influence thermalization, they also demonstrate that there is not a sharp distinction between random and quasiperiodic disorder in the prethermal regime.

	Based on the interpretation of the NEE phenomenology as belonging to the prethermal regime, we see little evidence for the presence of an asymptotic many-body mobility edge in past numerics~\cite{Li2015a,Hsu2018_PRL,Li2016,Tu2023b}. Rather, the various observations which led to this conjecture seem to be well explained by the maximum accessible timescale $\tau_\text{finite}$ crossing between the information spreading timescale $\tau_\text{info}$ and the particle (or magnetization) spreading timescale $\tau_\text{particle}$, as illustrated in Fig.~\ref{fig:fig1}.

	That we can see signatures of the NEE regime in autocorrelation functions, rather than past entanglement-based probes, indicates that this regime can in principle be observed in experiments~\cite{Schreiber2015,Choi2016,Luschen2017,Roushan2017,Luschen2018,Lukin2019probing,Rispoli2019,Guo2021,Leonard2023}. However, our dynamical probes are rather indirect, and likely difficult to use in an experimental setting. The proposal that the \(\tau\to 0\) feature of the stretched exponential fit is related to non-ergodicity is speculative, with a more conservative conclusion being merely that the model has additional features other than many-body resonances which control the decay of autocorrelation functions. This feature of the fit is also likely to be greatly affected by experimental noise. Thus, it would be desirable to have a direct and reliable measurement of \(\tau_\text{info}\) and \(\tau_\text{particle}\) to more directly test our proposed mechanism for the NEE regime. Due to our lack of direct probes, further investigation is still required to settle the nature of the NEE regime. Studies of the eigenstate properties of the random wells model are currently underway~\cite{YiTing_randw}, and preliminary results show that the behaviors of the GPD and random wells models appear almost identical to the machine learning model of Ref.~\cite{Hsu2018_PRL}.

	There may, in general, be several more time scales than \(\tau_\text{info}\) and \(\tau_\text{particle}\), associated to the spreading of different classes of operators. Even at the time scales accessible to current numerics and experiments, we suspect there can be multiple distinct intervening NEE regimes between apparent localization and clear thermalization. Verifying this conjecture also requires developing reliable ways of identifying these timescales directly.

	Asymptotically, we expect that the delocalization of any operator will generically cause the delocalization of all operators. A generic perturbation will couple the previously localized operators to the delocalized class, allowing them to spread. Thus, there should be no NEE regime, in the form of partial localization, in the thermodynamic limit. However, it is interesting to ask whether the NEE regime can be considered a distinct dynamical phase to MBL in a non-standard thermodynamic limit~\cite{Gopalakrishnan2019nonstandard,Morningstar2023Floquet}.

	We mention that the GPD model has been studied experimentally~\cite{An2021}, and it would be interesting to analyze this experiment in the context of our finding of prethermal NEE features in the GPD model. In fact, an interesting experiment would be to directly compare the prethermal dynamics of the GPD model with the random wells and $t_1$--$t_2$ models, since all three models are accessible in cold atom systems.  We predict that the noninteracting GPD model will be similar to the noninteracting $t_1$--$t_2$ model, with both manifesting SPMEs (with the random wells model having no SPME) whereas the interacting GPD model will manifest NEE features similar to the interacting random wells model with the interacting $t_1$--$t_2$ model showing no such prethermal NEE features. More broadly, the GPD model exhibits interesting dynamical features beyond the prethermal NEE phenomenology discussed here, including the enhancement of localization by interactions~\cite{LiGPD}.

	Our proposal also raises the theoretical question of whether there are other mechanisms by which the thermalization time scales for different classes of operators may separate. Any process which induces this feature should produce an NEE regime. Our work using the statistical Jacobi approximation in the interacting GPD model could be a guide for such future studies of prethermal quantum dynamics.

	\section*{Acknowledgements}	The authors thank C. Beveridge, A. Chandran, S. Garratt, Y.-T. Hsu, and D. Vu for useful discussions. This work is supported by the Laboratory for Physical Sciences. The authors acknowledge the University of Maryland supercomputing resources (\href{https://hpcc.umd.edu}{https://hpcc.umd.edu}) made available for conducting the research reported in this paper.

	\appendix

	\section{Eigenstate properties of the random wells model}\label{sec:randw_eigen}

	\begin{figure}[t]
		\includegraphics[trim=7 10 0 0, clip, scale=0.75]{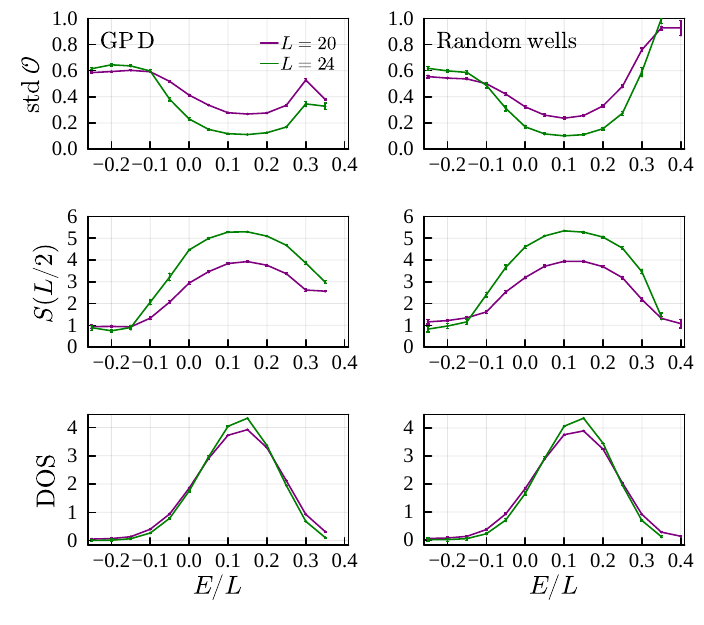}
		\caption{Comparison of the standard deviation of the observable $\mathcal{O}$ (Eq.~\eqref{eqn:halfchain}) among 100 consecutive eigenstates of energy close to \(E\), and the half-chain entanglement entropy between the GPD and the random wells models at $W=0.35$. Error bars indicate the $68\%$ bootstrap confidence interval due to the random choice of $\phi$ or disorder realizations. The almost quantitative agreement (except at the very right end of the spectrum) indicates that the two models are essentially indistinguishable at these system sizes. }
		\label{fig:entropy}
	\end{figure}

	In this appendix, we study the eigenstate properties of the random wells model using similar quantities to Ref.~\cite{Li2015a}, which were used to support the existence of an NEE regime in the GPD model.

	In Fig.~\ref{fig:entropy}, we show the standard deviation of the collection of eigenstate expectation values \(\{\mathcal{O}(E)\}\), defined by
	\begin{equation}\label{eqn:halfchain}
	  \mathcal{O}(E) = \sum_{j=1}^{L/2}\langle\psi_E|S^z_j|\psi_E\rangle,
	\end{equation}
	where $E$ runs over 100 consecutive energy eigenstates. In a system which satisfies the eigenstate thermalization hypothesis (ETH), this standard deviation should be very small, and become smaller with increasing system size~\cite{DAlessio2016}. We also show the average half-chain von Neumann entanglement entropy $S(L/2)$ of an eigenstate in that range. We compare these quantities between the interacting GPD and the random wells model, both at $W=0.35$, with parameters the same as in the main text.

	Except at very high energy densities, the two quantities (and the DOS of these two models) not only show qualitative similarity, but also agree almost quantitatively. This suggests that the GPD model is essentially indistinguishable from the random wells model at numerically tractable system sizes.

	In particular, the argument in Ref.~\cite{Li2015a} for the existence of an NEE regime based on similar numerical data carries through in exactly the same way for the random wells model. The decrease of $\operatorname{std}\mathcal{O}$ at low to middle energy densities---between \(E/L = -0.1\) and \(E/L=0.05\)---suggests that the system may be crossing from a non-ergodic regime (where consecutive eigenstates are very distinct, producing large fluctuations in $\mathcal{O}(E)$) to an ergodic regime (where consecutive eigenstates look similar). On the other hand, the entanglement entropy shows a change from area law ($S(L/2)\sim\text{const.}$) to approximately volume law ($S(L/2)\propto L$) at the slightly lower value $E/L\approx -0.15$, which suggests that the system goes from fully localized to extended at that energy density. The separation of these two crossovers implies that an NEE regime exists in both the interacting GPD and the random wells models at middle-low energy densities.

	The data presented in Fig.~\ref{fig:entropy} is not as clear as in Ref.~\cite{Li2015a} as we only use two different system sizes and work at smaller system sizes than their study. Note also that we use a higher filling fraction, among other parameter differences, compared to Ref.~\cite{Li2015a}. Ongoing study~\cite{YiTing_randw} shows that the random wells model also reproduces essentially the same data as in Ref.~\cite{Li2015a} using the setup of that work.

	\begin{figure}[b]
		\includegraphics[trim=3 0 0 0, clip, scale=0.75]{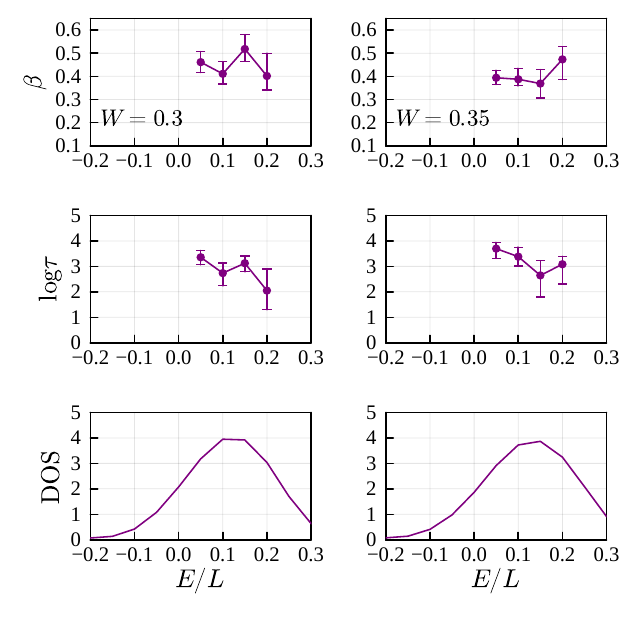}
		\caption{The fit parameters $\beta$ (top) and $\log\tau$ (middle) obtained from the shifted stretched exponential fit of $C(t)$, and the normalized density of states (bottom) for the $t_1$--$t_2$ model with GPD potential at $L=20$ and two selected values of $W$, each as a function of energy density. This model shows no signature of an NEE regime, consistent with the arguments of Sec.~\ref{sec:NEE}. (Compare to the GPD model in Fig.~\ref{fig:fits}.) }
		\label{fig:t1t2GPD}
	\end{figure}

	The difference in the position of the NEE regime as diagnosed by the dynamical properties presented in the main text (at middle-high energy densities) and the eigenstate properties studied here (at middle-low energy densities) is expected due to the difference in the probing time scale $\tau_\text{finite}$. Since eigenstate properties probe the longest possible timescale at a given system size, the $\tau_\text{finite}$ here is longer than the dynamical study in the main text. From Fig.~\ref{fig:fig1}c, we see that the parameter regime which shows the MBL behavior will drift to the NEE behavior, and the one that shows the NEE behavior will drift to the ETH behavior, upon increasing $\tau_\text{finite}$. Therefore, in our case, the NEE regime is expected to drift towards lower energy densities when using a larger $\tau_\text{finite}$.

	\section{The \texorpdfstring{$t_{1}$--$t_{2}$}{t1-t2} model with the GPD potential}\label{sec:t1t2GPD}

	In this appendix, we provide another verification of our proposed mechanism for the NEE regime described in Sec.~\ref{sec:NEE} by adding next-nearest-neighbor hopping to a model with deep wells. Our proposed mechanism predicts that this addition should remove the NEE regime.

	The model is defined by replacing the cosine term in Eq.~(\ref{eq:t1t2}) with the GPD potential in Eq.~(\ref{eq:GPD}), with the same parameters as in the main text. This is equivalent to adding next-nearest-neighbor hopping to the GPD results presented in Sec.~\ref{sec:AA_GPD}. The next-nearest-neighbor hopping avoids the large detuning of the deep well, making the NEE regime disappear. The results shown in Fig.~\ref{fig:t1t2GPD} support this picture, as there is no sudden decrease of $\tau$ at higher energy densities. This result supports our explanation of the NEE regime in Sec.~\ref{sec:NEE}.

	\bibliographystyle{apsrev4-2}
	\bibliography{references}

\end{document}